\documentclass[a4paper]{article}
\usepackage{xcolor}
\usepackage{INTERSPEECH2022}
\usepackage{bbding}
\usepackage{hyperref}

\title{Pre-Training Transformer Decoder for End-to-End ASR Model with \ Unpaired Speech Data}

\name{Junyi Ao$^{1,\dag}$, Ziqiang Zhang$^{2,\dag}$, Long Zhou$^3$, Shujie Liu$^3$, Haizhou Li$^1$, \\
Tom Ko, Lirong Dai$^{2}$, Jinyu Li$^3$, Yao Qian$^3$, Furu Wei$^3$\thanks{$^{\dag}$Equal contribution. Work done during internship at Microsoft Research Asia.}}
\address{
  $^1$School of Data Science, The Chinese University of Hong Kong (Shenzhen) \\
  $^2$NEL-SLIP, University of Science and Technology of China \\
  $^3$Microsoft
 }
\email{}

\begin{document}

\maketitle
\begin{abstract}
%When pre-training an encoder-decoder automatic speech recognition (ASR), typically we either pre-train the encoder with speech data or the decoder with text data. 
This paper studies a novel pre-training technique with unpaired speech data, Speech2C, for encoder-decoder based automatic speech recognition (ASR).
Within a multi-task learning framework, we introduce two pre-training tasks for the encoder-decoder network using acoustic units, i.e., pseudo codes, derived from an offline clustering model.
One is to predict the pseudo codes via masked language modeling in encoder output, like HuBERT model, while the other lets the decoder learn to reconstruct pseudo codes autoregressively instead of generating textual scripts. 
In this way, the decoder learns to reconstruct original speech information with codes before learning to generate correct text.
Comprehensive experiments on the LibriSpeech corpus show that the proposed Speech2C can relatively reduce the word error rate (WER) by 19.2\% over the method without decoder pre-training, and also outperforms significantly the state-of-the-art wav2vec 2.0 and HuBERT on fine-tuning subsets of 10h and 100h. 
We release our code and model at \url{https://github.com/microsoft/SpeechT5/tree/main/Speech2C}.
%Our proposed method also outperforms the state-of-the-art wav2vec2.0 and HuBERT model. 
%Furthermore, our method can also be used to continually pre-train based on a pre-trained speech encoder model.

\end{abstract}

\section{Introduction}

% Speech Pre-training
%Recently, self-supervised learning methods have been used to significantly improve the performance of natural language processing tasks, such as automatic speech recognition (ASR) \cite{baevski2019effectiveness,fan2019unsupervised,wang2020unsupervised,chung2020generative,wang2021unispeech,jiang2021further}.

Self-supervised learning has been shown effective in natural language processing (NLP), e.g., BERT \cite{devlin2018bert} and BART \cite{lewis2019bart}, where it makes use of a large amount of unlabeled data for pre-training to improve the performance of downstream tasks, such as automatic speech recognition (ASR) \cite{baevski2019effectiveness,fan2019unsupervised,wang2020unsupervised,chung2020generative,wang2021unispeech,jiang2021further}.

%Self-supervised learning methods are categorized into discriminative methods, generative methods, and multi-task learning methods. 
By their training objectives, self-supervised methods can be categorized into contrastive learning \cite{van2018representation,schneider2019wav2vec,baevski2020wav2vec} and reconstructive learning \cite{chung2020generative,hsu2021hubert,chen2021wavlm}.
In contrastive learning, CPC \cite{van2018representation} uses a probabilistic contrastive loss, which induces the latent space to capture information and uses an autoregressive model to classify future frames from negative examples.
Wav2vec \cite{schneider2019wav2vec} pre-trains a simple multi-layer convolutional neural network optimized via a noise contrastive binary classification task.
In reconstructive learning, APC \cite{chung2020generative} reconstructs the future frame with a unidirectional encoder via learning meaningful, non-specific, and transferable speech representations.
%Inspired by masked language modeling in BERT, 
% HuBERT \cite{hsu2021hubert} applies pseudo labels generated from an offline clustering step as the target to compute a BERT-like loss for model training, which can be boosted by an iterative process.
HuBERT \cite{hsu2021hubert} firstly generates pseudo labels from an offline clustering step, and they are used as the target to calculate a BERT-like loss for model training, which can be boosted by an iterative process.

% Drawback
% However, many previous studies only pre-train the speech encoder for various spoken downstream tasks, without pre-training the decoder for the end-to-end encoder-decoder based ASR models, which usually rely heavily on large amounts of high quality transcribed audio data.
However, many previous studies only pre-train the speech encoder for various spoken downstream tasks, while the decoder is not pre-trained for the end-to-end encoder-decoder based ASR models, which usually rely heavily on a large amount of transcribed audio data.
%Current industrial end-to-end ASR systems rely heavily on large amount of high quality transcribed audio data. However, transcribed data take substantial effort to obtain in industrial applications, while at the same time a lot of untranscribed data exist in online systems and cost little to collect. It is worthwhile to explore how to effectively use un-transcribed data to improve the performance of speech recognition systems when labeled data are limited.
Although there are some attempts to pre-train a Transformer decoder for end-to-end ASR model, they have to use additional unpaired text data \cite{fan2019unsupervised,gao2021pre,ao2021speecht5}.
%This article attempts to answer a question:
\textit{Can we pre-train the ASR decoder with speech-only data?}
There are two main challenges, (1) speech signals are continuous and textual representations are discrete, and (2) the decoder is responsible for generating text which is very different from conventional speech representation (e.g., waveform, log-mel fbank feature, or hidden states).

% Our method

To address this problem, we propose a speech to code pre-trained model (\textbf{Speech2C}), trying to pre-train the encoder-decoder model with speech-only data.
Under the multi-task learning framework, we employ two pre-training tasks for encoder-decoder pre-training using speech-only data with the acoustic units learned from an offline clustering model, aka pseudo codes.
The first task predicts the pseudo codes via masked language modeling (MLM) in encoder output, like HuBERT model.
% In addition to MLM loss, the decoder of Speech2C learns to reconstruct pseudo codes auto-regressively, instead of generating real text transcription, both of which are discrete representations and have some semantic information corresponding to the speech signal.
For the second one, the decoder of Speech2C learns to reconstruct reduced pseudo codes autoregressively, instead of generating real text transcription.
% Both text and codes are discrete representations and have some semantic information corresponding to the speech signal.
% With the pre-training methods, the decoder can learn how to reconstruct semantic information of speech with codes, which helps generate correct transcriptions.
% Moreover, the codes may highly correlate with the text of the speech according to our observation,
% which implies the decoder can implicitly learn text prediction in our pre-training framework.
Both codes and text are discrete representations and contain semantic information of the speech signal. 
According to our observation, a certain amount of codes are highly correlated with the transcriptions,
which implies that the decoder can learn text prediction in our pre-training framework.

We conduct massive experiments on the Librispeech dataset to validate the proposed Speech2C. 
To the best of our knowledge, this is the first work to pre-train an encoder-decoder model for ASR with only speech-only data, and our proposed Speech2C achieves a new state-of-the-art performance on the test set of Librispeech.
%can achieve a 19.2\% WERR compared with the model without decoder pre-training.

%The rest of this paper is organized as follows. In Section 2, we discuss the related work. Section 3 illustrates our proposed models Speech2C, including model architecture and two pre-training tasks.
%We share the experimental setting, results, and analyses in Section 4. Section 5 concludes the study.
%\textcolor{red}{to Junyi: please give the method a name. please remove the experiments from introduction, but rather add a paragraph - the rest of this paper is organized as folows. In Section 2, we discuss the related work. In Section 3, ...  -Haizhou}  

\begin{figure*}[!htp]
  \centering
  \includegraphics[width=13cm]{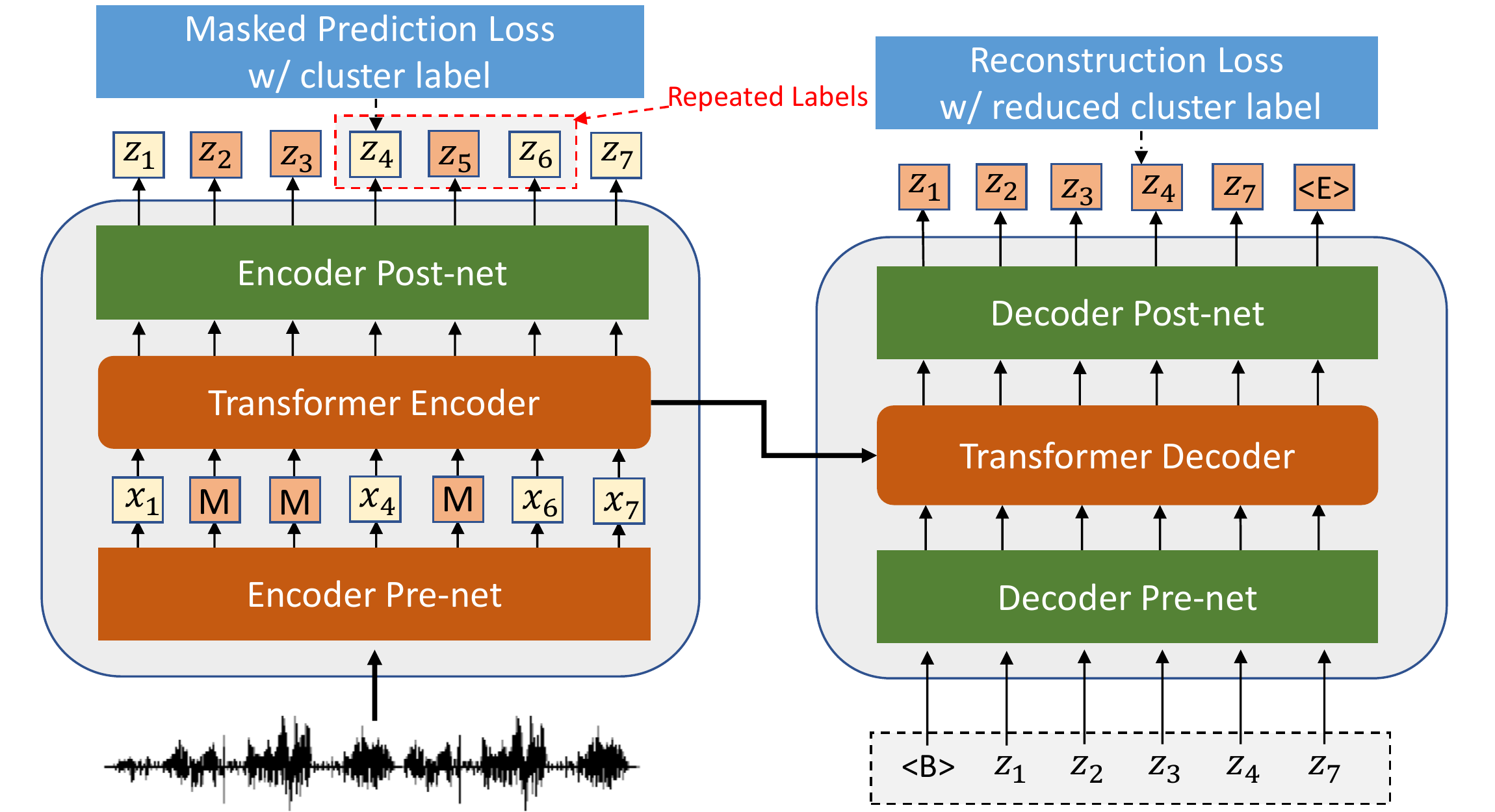}
  \caption{The framework of our proposed Speech2C model, which is pre-trained with masked prediction loss and reconstruction loss on unpaired speech data for end-to-end ASR. In this example, $z_4$, $z_5$ and $z_6$ are from the same class and can be reduced to a single label for the decoder. In the fine-tuning stage, we initialize the ASR model with pre-trained Speech2C by removing the encoder post-net and the decoder pre-net/post-net, since they are trained to process pseudo codes.}
  \label{fig_Speech2C}
 \vspace{-15pt}
\end{figure*}

\section{Related Work}

We consider our work most related to HuBERT \cite{hsu2021hubert}, which benefits from an offline clustering step to provide pseudo labels for a BERT-like pre-training.
The backbone of HuBERT includes a convolutional feature encoder and a Transformer context encoder.
During pre-training, HuBERT first uses k-means to learn the initial quantizer that maps speech signals to discrete labels, and performs BERT-style pre-training where the inputs are masked speech signals and prediction targets are discrete labels. 
Moreover, HuBERT allows refinement on the pseudo label by further using the pre-trained model as the new quantizer to train a new iteration of the model. 
%It repeats the process to improve the pre-training results iteratively.
However, HuBERT model only pre-trains a speech encoder, leaving the decoder not pre-trained for the encoder-decoder based tasks, such as end-to-end ASR \cite{li2021recent}.
Based on HuBERT encoder, our proposed Speech2C model can also pre-train a Transformer decoder with pseudo label from the clustering model. 
%Besides, some research works \cite{lakhotia2021generative,lee2021direct} also remove sequential repetitions of codes, but they mainly focus on reconstructing audios from the reduced codes, which are designed for the text to speech (TTS) and speech to speech translation tasks.
Besides, some research works \cite{lakhotia2021generative,lee2021direct} attempt to leverage pseudo codes to reconstruct audios, but they are mainly designed for speech resynthesis and speech to speech translation tasks.

In addition, an idea was explored to pre-train a decoder for end-to-end ASR \cite{fan2019unsupervised,gao2021pre,ao2021speecht5}. 
The authors in \cite{fan2019unsupervised} employ a single speaker text to speech (TTS) system to generate synthesized speech from a large number of transcripts, and use the generated speech-text pairs to pre-train the decoder.
In \cite{gao2021pre}, unpaired text data are used to pre-train the transformer decoder, which is pre-trained as a conditional language model by constructing empty or artificial states to replace the real encoder hidden states.
Leveraging large-scale unpaired speech and text data, SpeechT5 \cite{ao2021speecht5} pre-trains a shared encoder-decoder model for various spoken language tasks.
However, all previous work still utilize text data to pre-train a decoder for end-to-end ASR.
In contrast, Speech2C is the first work to pre-train a Transformer decoder for ASR without text data.

% Self-supervised learning from speech consists in resolving pseudo-tasks not requiring human annotations as a pre-training to the real tasks to solve.

% Recently, self-supervised ASR pre-training achieved impressive performance, especially with minimal amounts of labeled
% data

% Masked reconstruction has also been widely investigated which masks part of the input and learn to reconstruct it.
% \section{Method}

%We propose a SpeechT5-S2T, an encoder-decoder based self-supervised speech pre-training for speech to text tasks. 

\section{Methods}

In this section, we first illustrate our model architecture, based on which, we present two pre-training tasks in our proposed Speech2C, the masked prediction loss for the encoder and the pseudo-code reconstruction loss for the decoder.

%In this paper, we study a pre-training method that utilizes unpaired speech data to pre-train the Transformer decoder. 
%During the pre-train process, the decoder is pre-trained as a language model with empty encoder states to skip the src MHA component. 
%We further force the decoder to predict the next token with artifical states rather than the real encoder states, to train the parameters of src MHA. 
%The decoder is pre-trained to predict the pseudo codes from left to right.
%We believe this operation can help the decoder learn how to generate text sequences.
%Next, we present each component in Speech2C, following with two pre-training tasks.

\subsection{Model Architecture}

The model architecture of the proposed Speech2C for pre-training is composed of an encoder network that extracts latent speech representations from raw acoustic
inputs and learns contextualized speech representations, and a decoder network for autoregressively reconstructing the pseudo codes of corresponding source speech.
Both encoder network and decoder network are boosted with relative positional encoding \cite{shaw-etal-2018-self}.
%We employ the relative positional encoding \cite{shaw-etal-2018-self} for both of the encoder and decoder network.
Figure \ref{fig_Speech2C} illustrates the framework of the proposed Speech2C.

%\subsubsection{Encoder Network}

The encoder network follows the HuBERT \textsc{Base} architecture, with an encoder pre-net, the Transformer encoder, an encoder post-net, as shown on the left panel of Figure \ref{fig_Speech2C}.
More specifically, the encoder pre-net is a convolutional network for pre-processing waveform, which is  of seven 512-channel layers with strides [5,2,2,2,2,2,2] and kernel widths [10,3,3,3,3,2,2]. 
% The Transformer encoder consists of a stack of blocks, each of which comprises two components: a self-attention layer and a feed-forward network (FFN) following the self-attention layer.
The Transformer encoder contains 12 layers with model dimension 768, inner dimension 3072 and 12 attention heads.
Moreover, the encoder-post network contains a projection layer and a code embedding layer, which are utilized to convert hidden states into pseudo codes.

%\subsubsection{Decoder Network}
The decoder network also contains a decoder pre-net, the Transformer decoder, and a decoder post-net.
The pre-net transforms a code index into an embedding vector. 
% The Transformer decoder has a similar architecture to the Transformer encoder except that it includes a cross-attention mechanism after each self-attention layer that attends to the output of the encoder, and an autoregressive or causal self-attention is used to only attend to the past outputs.
The Transformer decoder has a similar architecture to the Transformer encoder except for the cross-attention and the masked self attention. 
Finally, the post-net transforms the hidden state into the probability distribution of codes, normalized by the softmax function.

\subsection{Self-Supervised Speech Pre-Training}
The proposed pre-training learning method for Speech2C has access to speech-only data.
%During pre-training only unlabeled speech data is used to. 
We introduce two pre-training tasks to pre-train the encoder-decoder model, including masked prediction loss for the encoder and reconstruction loss for the decoder.

\subsubsection{Masked prediction loss}

During pre-training, the encoder pre-net first generates a feature sequence $\mathbf{x}$ from waveform by down-sampling, then the Transformer encoder consumes masked acoustic features $\mathbf{\tilde{x}}$ and output hidden states $\mathbf{h}^L$.
%A masked prediction model f takes as inputand predicts a distribution over the target indeces at each timestep
Furthermore, the network including encoder post-net is optimized to predict the discrete target sequence $\mathbf{z}$, where each $z_t \in [C]$ is a $C$-class categorical variable. The distribution over codewords is parameterized with: 
\begin{equation}
    p_f(c|\mathbf{\tilde{x}}, t) = \frac{{\rm exp}({\rm sim}(\mathbf{h}_t^L\mathbf{W}, \mathbf{e}_c)/\tau)}{\sum_{c'=1}^{C}{\rm exp}({\rm sim}(\mathbf{h}_t^L\mathbf{W}, \mathbf{e}_{c'})/\tau)}
\label{eq1}
\end{equation} 
where $\mathbf{W}$ is a projection matrix, $\mathbf{h}_t^L$ is the output hidden state for step $t$ and layer $L$, $\mathbf{e}_c$ is the embedding for codeword $c$, sim($a, b$) computes the cosine similarity between two vectors and $\tau=0.1$ is used to scale the logit.
%Specifically, following HuBERT, we apply span mask strategies to the feature sequence $\mathbf{x}$ from encoder pre-net, where 8\% of timesteps are randomlyselected as start indices, and spans of 10 steps are masked.

Specifically, $\mathbf{\tilde{x}}$ comes from $\mathbf{x}$ by span mask strategies, where 8\% of timesteps are randomly selected as start indices, and spans of 10 steps are masked.
Based on the above distribution, we denote the cross-entropy loss computed over masked timesteps as
\begin{equation}
%\begin{aligned}
    \mathcal{L}_{mlm} = \sum_{t\in \mathcal{M}} \log p_f (\mathbf{z}_t|\mathbf{\tilde{x}}, t),
%\end{aligned}
\end{equation}
where, $\mathcal{M}$ denotes the set of masked timesteps, and $\mathbf{z}_t$ denotes the frame-level target at timestep $t$ from $\mathbf{Z}$.

\subsubsection{Pseudo-code reconstruction loss}

In addition to masked prediction loss in the encoder, we also design a reconstruction loss to pre-train the Transformer decoder.
Following the denoising autoencoder in BART \cite{lewis2019bart}, the decoder network is optimized to generate the reduced pseudo codes with the maximum likelihood estimation as
\begin{equation}
    \mathcal{L}_{mle} = \sum_{n =1}^{N} \log p (\mathbf{z}_n|\mathbf{z}_{<n},\mathbf{\tilde{x}}),
\end{equation}
where $N$ is the length of pseudo codes.

Pseudo codes are similar to real texts because (1) they are discrete representations and have fixed vocabulary; (2) they all have rich semantic information that can be aligned to speech fragments.
Hence, we believe this pre-training method on pseudo codes can help the decoder learn how to generate text sequences.
The pseudo codes of adjacent speech frames have some repeated codes, which may represent similar semantic information, but repeated words are rarely used consecutively in textual languages. 
To reduce the gap between pseudo code and real text, we remove the repeating code of adjacent speech frame, which will be studied in the ablation study.

\section{Experiments}

%\subsection{Dataset}

\subsection{Training Details}

All models are implemented in Fairseq{\footnote[1]{https://github.com/pytorch/fairseq}} \cite{ott2019fairseq}.
%The total number of model parameters is less than 152M.
For speech pre-training, we use the full 960 hours of LibriSpeech \cite{panayotov2015librispeech} audio without transcription.
We optimize the model with Adam \cite{kingma2014adam} by warming up the learning rate for the first 8\% of updates to a peak of $2 \times 10 ^{-4}$, which is linearly decayed for the following updates.
We pre-train the proposed Speech2C model on 32 V100 GPUs with a batch size of around 87.5s samples per GPU for 400k steps.
% We set the mask span to 10, and select 8\% of outputs from the encoder pre-net as the start indices.

\begin{table}[!h]
\begin{center}
\caption{\label{exp_100h} WER on the LibriSpeech test sets when training on the 10 hours and 100 hours subset. $\dagger$ indicates that the results are not reported in \cite{hsu2021hubert} and obtained by fine-tuning the pulic released model.}
% \textcolor{red}{to Junyi: please indicate WER \% for all numbers in the tables.}

\begin{tabular}{lccc}
\toprule
Model & LM  & test-clean & test-other \\
\midrule
\midrule
\textbf{\textit{10 hours subset}} \\
\hline
wav2vec2.0 \textsc{Base} \cite{baevski2020wav2vec} &  None  & 11.1 & 17.6 \\
HuBERT \textsc{Base} $\dagger$ \cite{hsu2021hubert}  & None  & 10.1 & 16.8 \\
% wav2vec2.0   \textsc{Large}  \cite{baevski2020wav2vec} & None  & 8.0 & 12.1 \\
% SpeechT5 \cite{ao2021speecht5} & w/o & 4.4 & 10.4 \\
% Baseline \cite{ao2021speecht5} & w/o  & 5.0 & 11.9 \\
%HuBERT   + Decoder CE Loss 8/4 & 104M & 4.8 & 10.7 & 4.9 & 10.9 \\
%HuBERT   + Decoder CE Loss 10/2 & 100M & 4.5 & 9.9 & 4.6 & 9.8 \\
Our Speech2C & None  & \textbf{7.8} & \textbf{13.1} \\
\midrule
%100hr ft (w/ LM) & \# Parameters & dev-clean & dev-other & test-clean & test-other \\
% Wav2vec2.0   Base & 95M & 2.9 & 7.4 & 3.2 & 7.8 \\
% Wav2vec2.0   Large & 317M & 2.9 & 5.7 & 3.2 & 6.1 \\
wav2vec2.0   \textsc{Base} \cite{baevski2020wav2vec} & 4-gram  & 4.3 & 9.5 \\
wav2vec2.0   \textsc{Base} \cite{baevski2020wav2vec} & Transf  & 3.2 & 7.8 \\
HuBERT \textsc{Base} \cite{hsu2021hubert} & 4-gram  & 4.3 & 9.4 \\
% wav2vec2.0   \textsc{Large} \cite{baevski2020wav2vec} & Transf  & 3.2 & 6.1 \\
%Wav2vec2.0   Large \cite{baevski2020wav2vec} & w/ & 317M & 2.1 & 4.8 & 2.3 & 5.0 \\
% SpeechT5 \cite{ao2021speecht5} & w/  & 2.4 & 5.8 \\
%HuBERT   + Decoder CE Loss 8/4 & 104M & 3.3 & 7.3 & 3.4 & 7.7 \\
%HuBERT   + Decoder CE Loss 10/2 & 100M & 3.0 & 6.8 & 3.2 & 7.2 \\
% HuBERT   + Decoder CE Loss 12/6 & 152M & 3.1 & 6.5 & 3.1 & 7.0 \\
% Baseline \cite{ao2021speecht5} & w/  & 2.5 & 6.3 \\
Our Speech2C & Transf  & \textbf{3.1} & \textbf{7.0} \\
\midrule
\midrule
\textbf{\textit{100 hours subset}} \\
\hline
wav2vec2.0 \textsc{Base} \cite{baevski2020wav2vec} & None  & 6.1 & 13.3 \\
wav2vec2.0 \textsc{Large} \cite{baevski2020wav2vec} & None  & 4.7 & 9.0 \\
HuBERT \textsc{Base} $\dagger$ \cite{hsu2021hubert}  & None  & 6.3 & 13.2 \\
SpeechT5 \cite{ao2021speecht5} & None & 4.4 & 10.4 \\
Baseline & None  & 5.0 & 11.9 \\
%HuBERT   + Decoder CE Loss 8/4 & 104M & 4.8 & 10.7 & 4.9 & 10.9 \\
%HuBERT   + Decoder CE Loss 10/2 & 100M & 4.5 & 9.9 & 4.6 & 9.8 \\
Our Speech2C & None & \textbf{4.3} & \textbf{9.0} \\
\midrule
%100hr ft (w/ LM) & \# Parameters & dev-clean & dev-other & test-clean & test-other \\
% Wav2vec2.0   Base & 95M & 2.9 & 7.4 & 3.2 & 7.8 \\
% Wav2vec2.0   Large & 317M & 2.9 & 5.7 & 3.2 & 6.1 \\
wav2vec2.0   \textsc{Base} \cite{baevski2020wav2vec} & 4-gram  & 3.4 & 8.0 \\
wav2vec2.0   \textsc{Base} \cite{baevski2020wav2vec} & Transf  & 2.6 & 6.3 \\
%Wav2vec2.0   Large \cite{baevski2020wav2vec} & w/ & 317M & 2.1 & 4.8 & 2.3 & 5.0 \\
HuBERT \textsc{Base} \cite{hsu2021hubert} & 4-gram  & 3.4 & 8.1 \\
SpeechT5 \cite{ao2021speecht5} & Transf  & 2.4 & 5.8 \\
%HuBERT   + Decoder CE Loss 8/4 & 104M & 3.3 & 7.3 & 3.4 & 7.7 \\
%HuBERT   + Decoder CE Loss 10/2 & 100M & 3.0 & 6.8 & 3.2 & 7.2 \\
% HuBERT   + Decoder CE Loss 12/6 & 152M & 3.1 & 6.5 & 3.1 & 7.0 \\
Baseline & Transf  & 2.5 & 6.3 \\
Our Speech2C & Transf  & \textbf{2.4} & \textbf{5.2} \\
\bottomrule
\end{tabular}
\end{center}
\vspace{-24pt}
\end{table}
%All of the experiments were based on the LibriSpeech \cite{panayotov2015librispeech} corpus.
For the fine-tuning, we employ 10 hours and 100 hours as the supervised paired corpus, and use the character set as the model units for the text.
Because pseudo codes and textual characters have different vocabulary, we initialize the ASR model with pre-trained Speech2C without the encoder post-net and decoder pre-net/post-net, which are trained from scratch in our fine-tuning model.
We utilize the CTC and cross-entropy loss to fine-tune the model \cite{shinji2017hybrid}, where the loss weights are 0.5 for both of them.
The models are trained on 16 v100 GPUs with a batch size of 100s samples per GPU.
The learning rate is warmed up for the first 10\% steps, held as a constant for the following 40\% steps, and is decayed linearly for the rest steps.
For the 10/100 hours subset, we train the model with a learning rate of 2e-5/4e-5 for 25k/80k, and fix the encoder part for the first 10k/25k steps.

% For the  hours subset, we train the model with a learning rate of 4e-5 for 80k, and fix the transformer encoder and the encoder pre-net for the first 25k steps.
For ASR inference, we apply the joint CTC/attention decoding \cite{hori-etal-2017-joint} and train a transformer language model (LM) by LibriSpeech-LM Corpus with the same architecture as in \cite{ao2021speecht5} for the shallow fusion \cite{gulcehre2015using} to boost the performance of our baseline and Speech2C.
We sweep over the weights of the language model and CTC from 0 to 1 on the dev-other subset and choose the best weights according to the WER.

\subsection{Main Results}

%\subsubsection{Results of Base Model}

The results of ASR on the 10 hours and 100 hours set of LibriSpeech are reported in Table \ref{exp_100h}. 
The WER is evaluated on the standard Librispech test-clean/other sets.
We compare with several state-of-the-art self-supervised approaches, including encoder-based wav2vec 2.0 \cite{baevski2020wav2vec} and HuBERT \cite{hsu2021hubert}, and encoder-decoder based SpeechT5 \cite{ao2021speecht5}, which utilizes the unpaired speech and text corpus to pre-train.
We build a strong encoder-decoder based ASR baseline system, which has the same model structure as our Speech2C, while the weights of the encoder are initialized by the HuBERT \textsc{Base} model.

Without LM fusion, the baseline outperforms wav2vec 2.0 \textsc{Base} and HuBERT \textsc{Base} with the help of the joint CTC/attention decoding, which shows the importance of the decoder.
Our proposed Speech2C model achieves significant improvements on all settings compared to wav2vec 2.0 \textsc{Base}, HuBERT \textsc{Base}, SpeechT5 \cite{ao2021speecht5} and our strong baselines, demonstrating the superiority of the proposed pre-training method.
% More specifically, our Speech2C gets a relative 18.7\% WER reduction on the average of all sets compared to baseline system, achieves the state-of-the-art performance.
More specifically, our Speech2C without LM gets a relative 19.2\% WER reduction on the average of all sets compared to the baseline system for 100h subset, which achieves state-of-the-art performance.
Furthermore, when decoding with LM shallow fusion, our Speech2C still obtains the lower WERs than the strong baseline on all sets.

%\subsubsection{Results of Large Model}

% \begin{table*}[!h]
% \begin{center}
% \caption{\label{exp_10h} Results of ASR on the LibriSpeech dev and test sets when training on the 10 hours subset of LibriSpeech. }
% \begin{tabular}{c|ccccc}
% \hline
% 10hr ft (w/o LM) & \# Parameters & dev-clean & dev-other & test-clean & test-other \\
% \hline
% Wav2vec2.0   Base & 95M & 10.9 & 17.4 & 11.1 & 17.6 \\
% Wav2vec2.0   Large & 317M & 8.1 & 12.0 & 8.0 & 12.1 \\
% HuBERT   with Rel Pos Enc & 95M & 8.8 & 14.9 & 8.7 & 15.2 \\
% %HuBERT   + Decoder CE Loss 8/4 & 104M & 9.7 & 15.5 & 10.0 & 15.9 \\
% %HuBERT   + Decoder CE Loss 10/2 & 100M & 8.4 & 13.5 & 8.5 & 14.1 \\
% Our Speech2C & 152M & 7.9 & 12.8 & 7.8 & 13.1 \\
% \hline
% 10hr ft (w/ LM) & \# Parameters & dev-clean & dev-other & test-clean & test-other \\
% \hline
% Wav2vec2.0   Base & 95M & 2.9 & 7.4 & 3.2 & 7.8 \\
% Wav2vec2.0   Large & 317M & 2.9 & 5.7 & 3.2 & 6.1 \\
% %HuBERT   + Decoder CE Loss 8/4 & 104M & 3.3 & 7.3 & 3.4 & 7.7 \\
% %HuBERT   + Decoder CE Loss 10/2 & 100M & 3.0 & 6.8 & 3.2 & 7.2 \\
% Our Speech2C & 152M & 3.1 & 6.5 & 3.1 & 7.0 \\
% \hline
% \end{tabular}
% \end{center}
% \end{table*}

\subsection{Ablation Study}

We present a series of ablation studies in the following sections to learn how code reduction, continuing pre-training, and model layer numbers affect performance.
The models for ablation studies are pre-trained on 960 hours and fine-tuned on the 100-hour subset using fixed hyperparameters.

\subsubsection{Effect of code reduction}

Here, we start with probing the effectiveness of the k-means clustering algorithm concerning code reduction. 
In Table \ref{exp_repeated}, we summarize the results of Speech2C with repeated pseudo codes when calculating reconstruction loss, and the Speech2C with repeated codes performs slightly worse than Speech2C with reduced codes. 
Moreover, reducing repeated codes has the following advantages: (1) the average length of reduced codes 
%on dev-other subset
is significantly shorter than that of repeated codes, which will accelerate the training progress of Speech2C; (2) removing repeated codes does not loss semantic information, and brings pseudo codes closer to the text. 

\begin{table}[!h]
\begin{center}
\caption{\label{exp_repeated} Comparison of Speech2C with repeated codes or reduced codes in terms of average length and WER.}
\begin{tabular}{l|ccc}
    \toprule
    Model & Length & test-clean & test-other \\
    \midrule
    Speech2C (repeated) & 358  & 4.4 & 9.4 \\
    Speech2C (reduced) & 216 & 4.3 & 9.0 \\
    \bottomrule
\end{tabular}
\end{center}
\vspace{-25pt}
\end{table}

\subsubsection{Continued pre-training  from released model}

In addition to pre-train Speech2C from scratch, our method can also support continually pre-train from a pre-trained speech encoder model, such as pre-trained HuBERT.
Table \ref{exp_ct} shows the experimental results of two pre-trained Speech2C.
Although two pre-training methods achieve comparable performance, initializing the encoder of Speech2C with HuBERT can speed up the convergence and reduce the training time. 

\begin{table}[!h]
\begin{center}
\caption{\label{exp_ct} WER scores of Speech2C trained from scratch or initialized by HuBERT encoder.}
\begin{tabular}{l|cc}
    \toprule
    Model  & test-clean & test-other \\
    \midrule
     Speech2C (from scratch) & 4.3 & 9.0 \\
    Speech2C (from HuBERT) &  4.1 & 9.1 \\
    \bottomrule
\end{tabular}
\end{center}
\vspace{-25pt}
\end{table}

\subsubsection{Effect of layer numbers}
Compared to wav2vec and HuBERT which adopt CTC inference based on speech encoder, the encoder-decoder based ASR (eg., SpeechT5 and Speech2C) adds an external decoder to generate text. 
To reduce the influence of model parameters, we reduce the model layers of encoder-decoder model to the similar parameters of encoder model. 
As shown in  table \ref{exp_layer}, we list the experimental results of Speech2C with different encoder and decoder layers.
We change the default setting of 12 encoder layers and 6 decoder layers to the total same layer number of wav2vec2.0 \textsc{Base} and HuBERT \textsc{Base}, such as 10 encoder layers and 2 decoder layers.
Results show that although reducing the model layers degrade the performance of Speech2C, it still behaves better than encoder based ASR model.
\begin{table}[!h]
\footnotesize
\begin{center}
\caption{\label{exp_layer} WER scores of Speech2C with different layer numbers.}
\begin{tabular}{l|cccc}
    \toprule
    Model & Size & Layer (enc-dec) & test-clean & test-other \\
    \midrule
    wav2vec2.0& 95M & 12-0 & 6.1 & 13.3 \\
    HuBERT& 95M & 12-0 & 6.3 & 13.2 \\
    SpeechT5& 154M & 12-6 & 4.4 & 10.4 \\ 
    \midrule   
    Speech2C & 104M & 8-4 & 4.9 & 10.9 \\
    Speech2C & 100M & 10-2 & 4.6 & 9.8 \\
    Speech2C & 152M & 12-6 & 4.3 & 9.0 \\
    % \midrule
    %  Speech2C (8-4) & & 2.4 & 5.8 \\
    % Speech2C (10-2) &  & 2.3 & 5.3 \\
    % Speech2C (12-6) &  & 2.4 & 5.2 \\
    \bottomrule
\end{tabular}
\end{center}
\vspace{-25pt}
\end{table}

\subsection{Analysis}
\begin{figure}[!htp]
  \centering
  \includegraphics[width=7.3cm]{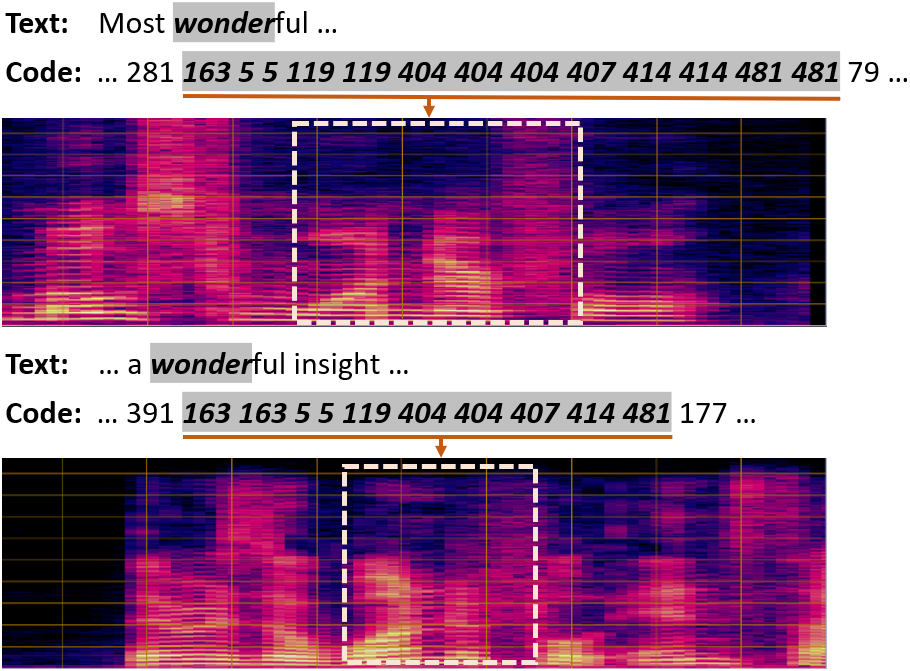}
  \caption{The transcripts, pseudo codes and Mel-Spectrum from two sentences, where the white boxes and the bold pseudo codes corresponding to the subword "wonder".}
  \label{fig_example}
 \vspace{-22pt}
\end{figure}

In this section, we want to answer a question: why pre-train the decoder with pseudo code can help text generation in ASR?
Frames of the same phonemes are more likely labeled as similar sequences of pseudo codes. 
We give an example chosen from the LibriSpeech dataset to show the high relevance of pseudo codes to text data and the patterns inherent in them.
As shown in Figure \ref{fig_example}, the codes corresponding to “wonder” in different samples are also similar and have an obvious pattern, which demonstrates that pseudo codes can be regarded as an intermediary language between waveform and transcription.
% \{\textcolor{red}{add a example from Junyi}\}
% Furthermore, pseudo codes can be regarded as an intermediary language, and pseudo codes and speech fragments have some alignment relationship.
\vspace{-7pt}

\section{Conclusion}

This paper proposes a novel model, Speech2C, to pre-train an encoder-decoder model with  speech-only data.
We present two pre-training tasks including masked prediction task and reconstruction task by taking advantage of acoustic units, i.e. pseudo codes derived from an offline clustering model.
Massive experiments and analyses on the LibriSpeech dataset show the effectiveness and superiority of our proposed Speech2C.
To the best of our knowledge, Speech2C is the first work to pre-train an encoder-decoder based ASR model with  speech-only data.
For future work, we will pre-train a multilingual Speech2C to address cross-lingual tasks, such as speech translation.

\section{Acknowledgements}
\label{sec:acknowledgements}

This research is supported by the internal project of Shenzhen Research Institute of Big Data under the Grant No. T00120220002, the Guangdong Provincial Key Laboratory of Big Data Computing under the Grant No. B10120210117-KP02, The Chinese University of Hong Kong, Shenzhen (CUHK-Shenzhen) and the University Development Fund, CUHK-Shenzhen, under the Grant No. UDF01002333 and UF02002333.
\bibliographystyle{IEEEtran}

\bibliography{mybib}

\end{document}